\documentstyle[11pt,preprint,tighten,graphicx,aps]{revtex}

\begin{document}

\def\slash#1{#1 \hskip -0.5em / } 
\def\beq{\begin{equation}}
\def\eeq{\end{equation}}
\def\beqy{\begin{eqnarray}}
\def\eeqy{\end{eqnarray}}

\thispagestyle{empty}
\preprint{JLAB-THY-02-19}
\title{$\rho$ Polarization and `Model Independent' Extraction of $|V_{ub}|/|V_{cd}|$ from
$D\to\rho\ell\nu$ and $B\to\rho\ell\nu$}
\author{
W. Roberts}
\address{Department of Physics, Old Dominion University, Norfolk, VA 23529,
USA \\
and \\
Continuous Electron Beam Accelerator Facility \\
12000 Jefferson Avenue, Newport News, VA 23606, USA.}

\maketitle
\begin{abstract}
We briefly discuss the predictions of the heavy quark effective theory for the semileptonic decays of a heavy
pseudoscalar to a light one, or to a light vector meson. We point out that measurement of combinations of differential
helicity decay rates at Cleo-c and the $B$ factories can provide a model independent means of extracting the ratio
 $|V_{ub}|/|V_{cd}|$. We briefly discuss the corrections to this prediction.
\end{abstract}

\section{Introduction and Motivation}
 
Extraction of the CKM matrix element $V_{ub}$ is one of the crucial ingredients
needed to determine the source of CP violation. A number of methods have been
suggested for extraction of this matrix element, both from inclusive and
exclusive decays. In this note, we point out that, modulo $1/m_c$, $1/m_b$ and
isospin corrections, the ratio $|V_{ub}|/|V_{cd}|$ can be determined using the
predictions of the heavy quark effective theory (HQET), and the measurement of
some polarization observables in the decays $D\to\rho\ell\nu$ and
$B\to\rho\ell\nu$, in a model independent way. 

Among the many measurements proposed for the proposed Cleo-c are precision
studies of the semileptonic decays $D\to \pi\ell\nu$ and $D\to\rho\ell\nu$: it is expected that the ratio 
$|V_{cd}|/|V_{cs}|$ can be measured to 1.3\%. In
particular, there should be ample statistics to study the polarization of the
vector mesons produced in the latter reaction. If this is the case, and if
similar measurements can be done at the $B$ factories, we propose measurements
that can provide the advertised ratio of CKM matrix elements with very small
theoretical errors.

In the next section we briefly discuss the predictions of HQET for these heavy
$\to$ light transitions, at leading order. Section III discusses a number of
experimental observables from these processes, and their dependence on the form
factors describing these semileptonic decays. In that section, we also present
the observables that allow extraction of the advertised ratio. In section IV we
present our conclusions.

\section{HQET And Form Factors}

The hadronic matrix elements for the decays $D\to \pi\ell\nu$ and 
$D\to\rho\ell\nu$ are
\begin{eqnarray}
\left<\pi(p^\prime)\left|\bar q\gamma_\mu c\right|D(p)\right>&=&f_+^{D\pi}(p+p^
\prime)_\mu+f_-^{D\pi}(p-p^\prime)_\mu,\nonumber\\
\left<\pi(p^\prime)\left|\bar q\gamma_\mu \gamma_5c\right|D(p)
\right>&=&0,\nonumber\\
\left<\rho(p^\prime,\epsilon)\left|\bar q\gamma_\mu c\right|D(p)
\right>&=&ig^{D\rho}\epsilon_{\mu\nu\alpha\beta}
\epsilon^{*\nu}(p+p^\prime)^\alpha(p-p^\prime)^\beta,\nonumber\\
\left<\rho(p^\prime,\epsilon)\left|\bar q\gamma_\mu\gamma_5 c
\right|D(p)\right>&=&f^{D\rho}\epsilon^*_\mu+a_+^{D\rho}\epsilon^*\cdot p(p+p^
\prime)_\mu
+a_-^{D\rho}\epsilon^*\cdot p(p-p^\prime)_\mu,
\end{eqnarray}
where the light quark $q$ can be either $u$ or $d$.
These decays are thus described in terms of six independent, {\it a 
priori} unknown form factors. The terms in $f_-$ and $a_-$
are unimportant when the lepton mass is ignored, since
\begin{equation}
(p-p^\prime)_\mu\bar\ell\gamma^\mu(1-\gamma_5)\nu_\ell =(k_\nu+k_
\ell)_\mu\bar\ell\gamma^\mu(1-\gamma_5)\nu_\ell
=m_\ell\bar\ell\gamma^\mu(1-\gamma_5)\nu_\ell.
\end{equation}
For the transitions from $B$ mesons, an analogous set of matrix elements and
form factors are necessary.
 
Using the Dirac matrix representation of heavy mesons, we may treat 
heavy-to-light transitions
using the same trace formalism that has been applied to 
heavy-to-heavy transitions \cite{hqet,falk}. In the effective theory, 
a heavy pseudoscalar meson ($D$ or $B$), denoted $P$,
traveling with velocity $v$ is represented as \cite{hqet,falk}
\begin{equation}
|{\cal P}(v)>=-\frac{1}{\sqrt{2}}\frac{1+\slash{v}}{2}\gamma_5\equiv {\cal M}_P(v),
\end{equation}
These states are normalized 
so that
\begin{equation}
\left<{\cal P}(v^\prime)\left|\right.{\cal P}(v)\right>=2v_0\delta^3
\left({\bf p-p^\prime}\right).
\end{equation}
The states of QCD and HQET are therefore related by
\begin{equation}
\left|P(v)\right>=\sqrt{m_P}\left|{\cal P}(v)\right>,
\end{equation}
where $m_P$ is the mass of the pseudoscalar meson. 
 
For the semileptonic transitions between such a heavy meson ($P$ 
meson) and a light pseudoscalar ($\pi$), the matrix element of 
interest is \cite{robertsledroit}
\begin{equation}
\left<\pi(p)\left|\bar q\Gamma h_v^{(Q)}\right|{\cal P}(v)
\right>={\rm Tr}\left[\gamma_5\left(\xi_1+\slash{p}\xi_2\right)\gamma_\mu 
{\cal M}_P(v)\right],
\end{equation}
where $h_v^{(Q)}$ is the heavy quark in the effective theory, and $q$ denotes a light quark ($u$ or $d$). For the transition to a light vector ($\rho$), the 
matrix elements are written
\begin{equation}
\left<\rho(p,\varepsilon)\left|\bar{q}\Gamma h_v^{(Q)}\right|{\cal P}(v)\right>={\rm Tr}
\left[\left\{\left(\xi_3+\slash{p}\xi_4\right)\varepsilon^*\cdot v+\slash{\varepsilon}^*
\left(\xi_5+\slash{p}\xi_6\right)
\right\}\Gamma{\cal M}_P(v)\right],
\end{equation}
where $\Gamma$ denotes either $\gamma_\mu$ or $\gamma_\mu\gamma_5$.

The form factors $\xi_i$ are independent of the mass of the heavy quark, and are therefore 
universal functions. Thus, they are valid for $D\to \pi (\rho)$ decays, as 
well as for $B\to \pi (\rho)$ decays.
This independence of the quark mass allows us to deduce, in a 
relatively straightforward manner, the scaling behavior of the usual 
form factors that
describe these transitions \cite{iw2}. More precisely, they allow us to relate the form factors for $D$
transitions to those for $B$ transitions. The relationships are
 
\begin{eqnarray} \label{ffmix}
f_+^B(v\cdot p)&=&\frac{1}{2}\left(\frac{m_B}{m_D}\right)^{1/2}
\left[f_+^D(v\cdot p)\left(1+\frac{m_D}{m_B}\right)
+f_-^D(v\cdot p)\left(\frac{m_D}{m_B}-1\right)\right],
\nonumber\\
f_-^B(v\cdot p)&=&\frac{1}{2}\left(\frac{m_B}{m_D}\right)^{1/2}
\left[f_-^D(v\cdot p)\left(1+\frac{m_D}{m_B}\right)
+f_+^D(v\cdot p)\left(\frac{m_D}{m_B}-1\right)\right],
\nonumber\\
f^B(v\cdot p)&=&\left(\frac{m_B}{m_D}\right)^{1/2}f^D(v\cdot p),
\nonumber\\
g^B(v\cdot p)&=&\left(\frac{m_D}{m_B}\right)^{1/2}g^D(v\cdot p),
\nonumber\\
a_+^B(v\cdot p)&=&\frac{1}{2}\left(\frac{m_D}{m_B}\right)^{1/2}
\left[a_+^D(v\cdot p)\left(1+\frac{m_D}{m_B}\right)
+a_-^D(v\cdot p)\left(\frac{m_D}{m_B}-1\right)\right],
\nonumber\\
a_-^B(v\cdot p)&=&\frac{1}{2}\left(\frac{m_D}{m_B}\right)^{1/2}
\left[a_-^D(v\cdot p)\left(1+\frac{m_D}{m_B}\right)
+a_+^D(v\cdot p)\left(\frac{m_D}{m_B}-1\right)\right],
\end{eqnarray}
where $f_+^D$ is the form factor appropriate to the $D\to \pi$ 
transition, while $f_+^B$ is the form factor appropriate to the $B\to 
\pi$
transition, and quantities on the left-hand-sides of eqns. 
(\ref{ffmix}) are evaluated at the same values of $v\cdot p$ as those 
on
the right-hand-sides. Omitted from each of eqn. (\ref{ffmix}) is a QCD 
scaling factor.
 
In the
limit of a heavy $b$ quark, the full current of QCD is replaced by 
\cite{qcd}
\begin{equation}
\bar q\Gamma b\to \bar q \Gamma h_v^{(b)}\left[\frac{\alpha_s(m_b)}{
\alpha_s(\mu)}\right]^{-\frac{6}{25}}.
\end{equation}
This arises from integrating out the $b$ quark, 
and matching the resulting effective theory onto full QCD at the
scale $m_b$, at one loop level. At the scale $m_c$, we must also 
integrate out the $c$ quark, but there is also the effect due to
running between $m_b$ and $m_c$. The net effect of this is that 
the form factors $\xi_i$ appropriate to the $b\to u$
transitions are related to those for the $c\to d$ transitions by
\begin{equation}
\xi_i^{b\to u}=\xi_i^{c\to d}\left[\frac{\alpha_s(m_b)}{
\alpha_s(m_c)}\right]^{-\frac{6}{25}}.
\end{equation}

\section{Observables}

In the differential decay rate of $D (B) \to\pi\ell\nu$, terms that depend on the form factor $f_-$ are proportional
to the mass of the lepton, and are thus very difficult to extract from experiment. Detecting the polarization of the
charged lepton offers the only possibility of extracting this form factor, but this is apparently a very remote prospect
with muons or electrons. Because of the predicted mixing of $f_+$ and $f_-$ in HQET, it becomes very difficult to
say anything about $B\to\pi$ observables based on information extracted from the corresponding $D\to\pi$ observables.
Assumptions about the form of $f_-$ can be and have been made, but this introduces some model dependence into any
information extracted. 

For the decay to the $\rho$, helicity amplitudes $H_\pm$ and $H_0$ can be defined as
\beqy
H_\pm(q^2)&=&f(q^2)\pm m_P k_\rho g(q^2),\nonumber\\
H_0(q^2)&=&\frac{1}{2m_\rho\sqrt{q^2}}\left[(m_P^2-m_\rho^2-q^2)f(q^2)+
4m_P^2k_\rho^2 a_+(q^2)\right],
\eeqy
where $k_\rho$ is the momentum of the daughter $\rho$ in the rest frame of the parent pseudoscalar. In terms of
these, the differential decay rate is written \cite{semileps}
\beqy\label{angles}
\frac{d\Gamma}{dq^2d\cos{\theta_\ell}d\cos{\theta_\rho}d\chi}&=&\frac{3G_F^2|V_{qQ}|^2k_\rho
q^2}{8(4\pi)^4m_P^2} \nonumber\\
&\times&\left\{\left[\left(1+\eta\cos{\theta_\ell}\right)^2\left|H_+(q^2)\right|^2+
\left(1-\eta\cos{\theta_\ell}\right)^2\left|H_-(q^2)\right|^2 \right]\sin^2{\theta_\rho}\right.\nonumber\\
&+&4\sin^2{\theta_\ell}\cos^2{\theta_\rho}\left|H_0(q^2)\right|^2-
2\sin^2{\theta_\ell}\sin^2{\theta_\rho}\cos{2\chi}H_+(q^2)H_-(q^2)\nonumber\\
&+&4\eta\sin{\theta_\ell}\sin{\theta_\rho}\cos{\theta_\rho}\cos{\chi}H_0(q^2)\nonumber\\
&\times&\left. \vphantom{\left|H_+(q^2)\right|^2}\left[
\left(1-\eta\cos{\theta_\ell}\right)H_-(q^2)-\left(1+\eta\cos{\theta_\ell}\right)H_+(q^2)\right]\right\},
\eeqy
where $\eta=+1$ for $B$ decays, and $-1$ for $D$ decays. The angles $\theta_\ell$, $\theta_\rho$ and $\chi$ are explained in
the figure.

With sufficient statistics in both $B$ and $D$ decays, this differential decay rate can provide means of extracting the ratio 
$|V_{ub}|/|V_{cd}|$. The differential decay rates into specific helicity states of the $\rho$ can be written as
\beq
\frac{d\Gamma_i}{dq^2}=\frac{G_F^2\left|V_qQ\right|^2}{96\pi^3} k_\rho\frac{q^2}{m_P^2}\left|H_i(q^2)\right|^2.
\eeq

Much of the difficulty of saying anything about the total rate in $B\to\rho\ell\nu$, based on measurements of 
$D\to\rho\ell\nu$, remains
because of the mixing of $a_+$ and $a_-$. As with the decays to pions, $a_-$ is, for the most part, only 
accessible through measurement of the polarization of the charged lepton. However, decays to transversely polarized
$\rho$ mesons are independent of this form factor, depending only on $f$ and $g$.

If $d\Gamma_\pm^D/dq^2$ can be measured or extracted at Cleo-c, and $d\Gamma_\pm^B/dq^2$ at $B$ factories, the
ratio of these two differential decay rates, at the same kinematic point $v\cdot p$ ($=E_\rho$ in the rest frame
of the parent hadron), depends only on known or measurable kinematic quantities, as the form factor dependence 
drops out at leading order in HQET. Actually, this is not quite true. This ratio will depend on the form factor 
ratio $r$ defined as
\begin{equation}
r(v\cdot p)=\frac{g^D(v \cdot p)}{f^D(v\cdot p)},
\end{equation}
but this ratio should be extractable from other observables in the $D\to\rho\ell\nu$ decay. More explicitly,
\begin{equation}
\frac{d\Gamma_\pm^B/dq^2}{d\Gamma_\pm^D/dq^2}=\frac{\left|V_{ub}\right|^2}{\left|V_{cd}\right|^2}
\frac{q_B^2m_D^2}{q_D^2m_B^2} \frac{\frac{m_B}{m_D}+
m_B m_D \left(k_\rho^B\right)^2 r^2(v\cdot p)\pm 2m_Bk_\rho^B r(v\cdot p)}
{1+m_D^2 \left(k_\rho^D\right)^2 r^2(v\cdot p)\pm 2m_Dk_\rho^D r(v\cdot p)}.
\end{equation}
Since the numerator and denominator of this ratio are evaluated at the same value of $E_\rho$, $k_\rho^D=k_\rho^B$. In
the expression above,
\beq
q_B^2=m_B^2+m_\rho^2-2m_B v\cdot p,\,\,\, q_D^2=m_D^2+m_\rho^2-2m_D v\cdot p.
\eeq

The combination $d\Gamma_+/dq^2+d\Gamma_-/dq^2$ is more easily accessible in these experiments (as the term proportional to
$(1+\cos^2{\theta_\ell})\sin^2{\theta_\rho}$ in the expression for the decay rate). The ratio of this quantity for $B$ decays to that 
for $D$ decays is
\beq
\frac{d\Gamma_+^B/dq^2+d\Gamma_-^B/dq^2}{d\Gamma_+^D/dq^2+d\Gamma_-^D/dq^2}=
\frac{\left|V_{ub}\right|^2}{\left|V_{cd}\right|^2}
\frac{q_B^2m_D^2}{q_D^2m_B^2} \frac{\frac{m_B}{m_D}+
m_B m_D \left(k_\rho^B\right)^2 r^2(v\cdot p)}
{1+m_D^2 \left(k_\rho^D\right)^2 r^2(v\cdot p)}.
\eeq
In addition, the differential rate $d\Gamma_{+-}/dq^2$, proportional to $H_+(q^2)H_-(q^2)$ may also be accessible (as the term
proportional to $\cos{2\chi}$ in the expression for the decay rate). If this is measured in $B$ and $D$ decays, the ratio is
\beq
\frac{d\Gamma_{+-}^B/dq^2}{d\Gamma_{+-}^D/dq^2}=\frac{\left|V_{ub}\right|^2}{\left|V_{cd}\right|^2}
\frac{q_B^2m_D^2}{q_D^2m_B^2} \frac{\frac{m_B}{m_D}-
m_B m_D \left(k_\rho^B\right)^2 r^2(v\cdot p)}
{1-m_D^2 \left(k_\rho^D\right)^2 r^2(v\cdot p)}.
\eeq
This ratio again depends only on known or measurable kinematic quantities, and the form factor ratio $r(v\cdot p)$.

Far more intriguing is the ratio of the difference of these differential helicity decay rates, related to the differential lepton
forward-backward asymmetry, and proportional to $\cos{\theta_\ell}\sin^2{\theta_\rho}$. This ratio takes the form
\begin{equation}
\frac{d\Gamma_+^B/dq^2-d\Gamma_-^B/dq^2}{d\Gamma_+^D/dq^2-d\Gamma_-^D/dq^2}
=-\frac{\left|V_{ub}\right|^2}{\left|V_{cd}\right|^2} \frac{q_B^2m_D}{q_D^2m_B},
\end{equation}
where a factor of $k_\rho^B/k_\rho^D$ has been set to unity, as explained above. Inclusion of the lowest order
radiative corrections means that the ratios above must be multiplied by the factor $\left[\frac{\alpha_s(m_b)}{
\alpha_s(m_c)}\right]^{-\frac{12}{25}}$.

At leading order, the last ratio of observables is completely independent of any form factor, and so should provide
a very good means of extracting the ratio $|V_{ub}|/|V_{cd}|$.

\section{Discussion and Conclusion}

We have suggested three measurements that can allow the extraction of $|V_{ub}|/|V_{cd}|$ in an
absolutely model independent manner. 
However, our results are subject to a number of corrections, which we discuss briefly here.

One correction that must be taken into account is isospin breaking. For the $D$ mesons, the
possible decay modes are $D^0\to\rho^-\ell^+\nu_\ell$ and $D^+\to\rho^0\ell^+\nu_\ell$, while
the corresponding $B$ decay modes are $\bar{B}^0\to\rho^+\ell\bar{\nu}_\ell$ and
$B^-\to\rho^0\ell\bar{\nu}_\ell$. The predictions given in the previous
section implicitly assume that the light component of the heavy meson is the same for both the
$D$ and $B$ decays. This means that comparisons can be made between the decays of the $D^0$ and
those of the $B^-$, or between the decays of the $D^+$ and those of the $\bar{B}^0$. In
each of these comparisons, the daughter $\rho$ is different, so some assumption of isospin
invariance among the form factors has to be made. Departures from this invariance can be expected to be
small. There is also an implicit assumption of isospin invariance in the discussion of the
radiative factor of $\left[\frac{\alpha_s(m_b)}{\alpha_s(m_c)}\right]^{-\frac{12}{25}}$.

By far the larger corrections are expected to come from the $1/m_c$ and $1/m_b$ contributions to
the matrix elements of interest. For decays to pions, the $1/m$ corrections require the introduction of eight new
universal form factors. The number of new form factors required for the decays to $\rho$ mesons 
will be twice that number.
For the decays to pions, Burdman and collaborators \cite{burdman} have used chiral symmetry arguments to obtain estimates of the normalizations of 
some of the new form factors at the non-recoil point, $v\cdot p=m_\pi$. It is not clear that analogous estimates can be obtained for
the $D\to\rho$ and $B\to\rho$ form factors.

The very nice result obtained in the previous section will undoubtedly be spoiled by $1/m$ 
corrections. Although there is no {\it a priori} reason to believe it, one can hope that such 
corrections are small. It is possible that, in the ratio of differential decay
rates, this turns out to be the case. This, however, is speculation. Nevertheless, perhaps the 
measurements suggested are of sufficient interest for
the $1/m$ corrections to be studied in detail in the near future.\\\\

The author thanks J. L. Goity for reading the manuscript, and for discussions.
This work was
supported by the National Science Foundation through grant \# PHY-9457892. This work was also supported 
by the Department of Energy through contract DE-AC05-84ER40150, under which the Southeastern Universities Research Association (SURA) operates the Thomas
Jefferson National Accelerator Facility (TJNAF), and through contract DE-FG05-94ER40832.

\begin{figure}
\caption{The angles of eqn. (\protect{\ref{angles}}). $\theta_\ell$ and $\theta_\rho$ are defined in the rest frames of the
lepton pair and the $\rho$, respectively. $\chi$ is the angle between the lepton plane and the hadron plane.}
\includegraphics[width=4.5in]{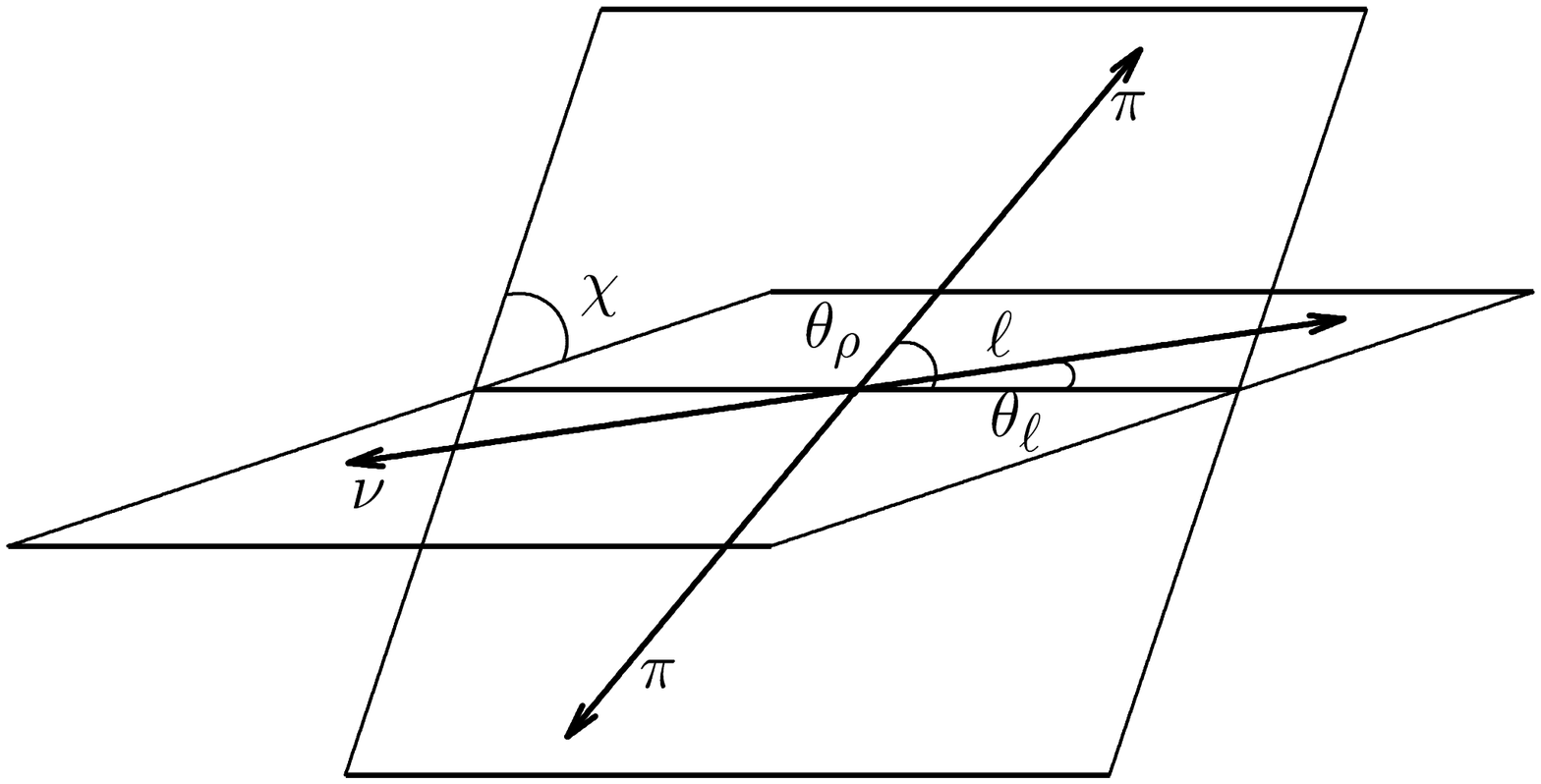}
\end{figure}
\end{document}